\documentclass[twocolumn,showpacs,superscriptaddress]{revtex4}

\usepackage[version=3]{mhchem} 
\usepackage{graphicx}

\usepackage{bm}         
\usepackage{amssymb}

\begin{document}

\title{Nanoscale multifunctional sensor formed by a Ni nanotube and a scanning Nb nanoSQUID}

\author{J.~Nagel}
\affiliation{%
  Physikalisches Institut and Center for Collective Quantum Phenomena in LISA$^+$,
  Universit\"at T\"ubingen,
  Auf der Morgenstelle 14,
  D-72076 T\"ubingen, Germany}
\author{A.~Buchter}
\affiliation{%
  Department of Physics, University of Basel, Klingelbergstrasse 82, CH-4056 Basel, Switzerland}
\author{F.~Xue}
\affiliation{%
  Department of Physics, University of Basel, Klingelbergstrasse 82, CH-4056 Basel, Switzerland}
\author{O.~F.~Kieler}
\author{T.~Weimann}
\author{J.~Kohlmann}
\author{A.~B.~Zorin}
\affiliation{%
  Fachbereich 2.4 "Quantenelektronik", Physikalisch-Technische Bundesanstalt, Bundesallee 100, 38116 Braunschweig, Germany}
\author{D.~R\"uffer}
\author{E.~Russo-Averchi}
\affiliation{%
  Laboratoire des Mat$\acute{e}$riaux Semiconducteurs, Institut des Mat$\acute{e}$riaux, Ecole Polytechnique F$\acute{e}$d$\acute{e}$rale de Lausanne, CH-1015 Lausanne, Switzerland}
\author{R.~Huber}
\author{P.~Berberich}
\affiliation{%
  Physik-Department E10, Technische Universit\"at M\"unchen, Garching, Germany}
\author{A.~Fontcuberta~i~Morral}
\affiliation{%
  Laboratoire des Mat$\acute{e}$riaux Semiconducteurs, Institut des Mat$\acute{e}$riaux, Ecole Polytechnique F$\acute{e}$d$\acute{e}$rale de Lausanne, CH-1015 Lausanne, Switzerland}
\author{D.~Grundler}
\affiliation{%
  Physik-Department E10, Technische Universit\"at M\"unchen, Garching, Germany}
\affiliation{%
  STI, Facult$\acute{e}$ Sciences et Technique de l'Ing$\acute{e}$nieur, Ecole Polytechnique F$\acute{e}$d$\acute{e}$rale de Lausanne, CH-1015 Lausanne, Switzerland}
\author{R.~Kleiner}
\author{D.~Koelle}
\email{koelle@uni-tuebingen.de}
\affiliation{%
  Physikalisches Institut and Center for Collective Quantum Phenomena in LISA$^+$,
  Universit\"at T\"ubingen,
  Auf der Morgenstelle 14,
  D-72076 T\"ubingen, Germany}
\author{M.~Poggio}
\affiliation{%
  Department of Physics, University of Basel, Klingelbergstrasse 82, CH-4056 Basel, Switzerland}
\author{M.~Kemmler}
\affiliation{%
  Physikalisches Institut and Center for Collective Quantum Phenomena in LISA$^+$,
  Universit\"at T\"ubingen,
  Auf der Morgenstelle 14,
  D-72076 T\"ubingen, Germany}
  
\date{\today}

\begin{abstract}
Nanoscale magnets might form the building blocks of next generation memories. To explore their functionality, magnetic sensing at the nanoscale is key. We present a multifunctional combination of a scanning nanometer-sized superconducting quantum interference device (nanoSQUID) and a Ni nanotube attached to an ultrasoft cantilever as a magnetic tip. We map out and analyze the magnetic coupling between the Ni tube and the Nb nanoSQUID, demonstrate imaging of an Abrikosov vortex trapped in the SQUID structure -- which is important in ruling out spurious magnetic signals -- and reveal the high potential of the nanoSQUID as an ultrasensitive displacement detector.  Our results open a new avenue for fundamental studies of nanoscale magnetism and superconductivity.
\end{abstract}

\pacs{%
85.25.Dq, 
74.78.Na, 
68.37.Rt 
75.75.+a 
85.25.CP, 
74.25.F- 
}

\maketitle

\section{Introduction}
There is a growing interest in the investigation of small spin systems, such as molecular magnets,\cite{Thomas96,Gatteschi03,Bogani08} single chain magnets,\cite{Bogani08a} single electrons,\cite{Bushev11} or cold atom clouds.\cite{Fortagh05}
Various detection schemes, e.g., magneto-optical spin detection,\cite{Maze08, Balasubramanian08} magnetic resonance force microscopy,\cite{Rugar04} or scanning-tunneling-microscopy assisted electron spin resonance,\cite{Manassen89,Durkan02} have been developed in order to detect such systems.
Unlike these techniques, superconducting quantum interference devices (SQUIDs) directly measure the stray magnetic flux produced by a small magnetic particle (SMP) with a large bandwidth.\cite{Wernsdorfer01,Gallop03}  This capability is especially interesting for the study of SMPs that support magnetic states not normally allowed in macroscopic magnets.\cite{Wang02, Topp08, Sreubel12, Streubel12a}

Here we couple a ferromagnetic Ni nanotube to a nanometer-scale SQUID (nanoSQUID), optimized for SMP detection.
%
%
A direct current (dc) SQUID is a superconducting loop, intersected by two Josephson junctions, and works as a flux-to-voltage transducer, i.e., the magnetic flux $\Phi$ threading the loop is converted into a voltage $V$ across the junctions, yielding a periodic $V(\Phi)$ characteristics with a period of one magnetic flux quantum $\Phi_0=h/2e$ (see, e.g., the review Ref.~\onlinecite{Kleiner04}).
Since the magnetic field distribution of a SMP is very close to that of a magnetic dipole, the figure of merit for such SQUIDs is the spin sensitivity $S_\mu^{1/2}=S_\Phi^{1/2}/\phi_\mu$, where $S_\Phi$ is the spectral density of flux noise power and $\phi_\mu\equiv\Phi/\mu$ is the coupling factor, i.e., the amount of magnetic flux coupled into the SQUID loop per magnetic moment $\mu\equiv|\vec{\mu}|$ of the SMP.
Both, $S_\Phi$ and $\phi_\mu$ can be optimized by scaling the SQUID down to nanometer dimensions.\cite{Foley09,Ketchen89,Bouchiat09,Nagel11}
Various fabrication techniques, e.g. electron-beam lithography,\cite{Voss80a,Nagel11a} focused ion beam milling,\cite{Troeman07,Hao08,Nagel11} atomic force microscopy anodization,\cite{Bouchiat01,Faucher09} self-aligned shadow evaporation,\cite{Finkler10} or a combination of electron-beam lithography with the use of carbon nanotube junctions,\cite{Cleuziou06} have been used to realize nanoSQUIDs.

The experimental determination of $S_\Phi$ poses no basic difficulties and was performed for various nanoSQUIDs that were fabricated, in some cases yielding very low values for $S_\Phi^{1/2}\approx 0.2-0.3\,\mu\Phi_0/\sqrt{{\rm Hz}}$.\cite{Voss80a,Hao08,Nagel11a}.
In contrast, the determination of $\phi_\mu$ is not straightforward, as it depends on the position $\vec{r}_p$ and orientation $\hat{e}_\mu$ of the magnetic moment of a SMP relative to the SQUID loop and on the geometry of the SQUID.
Up to now, $\phi_\mu$ has been determined only by numerical or analytical calculations, which often rely on strongly simplifying assumptions (e.g.~on the SQUID geometry or position and orientation of the SMP), restricting the validity of such approaches.\cite{Ketchen89,Bouchiat09, Granata09}
Recently, we proposed a routine for calculating $\phi_\mu(\hat{e}_\mu,\vec{r}_p)$, for a point-like particle in the 3-dimensional (3D) space above the SQUID loop.\cite{Nagel11,Nagel11a}
This routine takes explicitly into account the geometry in the plane of the SQUID loop, and is based on the numerical simulation of the 2-dimensional (2D) sheet current density in the SQUID loop, using London theory.\cite{Khapaev01a}
Still, this approach neglects the finite size of a SMP and the full 3D geometry of a SQUID.
Altogether, the determination of the spin sensitivity $S_\mu^{1/2}$ of nanoSQUIDs, as a figure of merit for SMP detection, requires the reliable determination of the coupling factor $\phi_\mu$.
Hence, the experimental determination of $\phi_\mu$ could serve as an important step forward, in order to test the theoretical approaches described above.

Ni nanotubes are widely investigated SMPs both experimentally \cite{Weber12, Rueffer12} and theoretically \cite{Escrig07, Escrig07a,Wang05a}. They are characterized by a well defined geometry due to the controlled production process and very small scale. When brought into a homogeneous magnetic field, defined magnetic states can be generated in the nanotube. A high enough axial field saturates the magnetization along the nanotube's long axis thus leading to a nearly dipolar magnetic field distribution.

Here, we present a multifunctional sensor system, consisting of a combination of a low-temperature magnetic force microscope (LTMFM) with a nanoSQUID.
This system allows for magnetization measurements of nanoscaled magnetic samples using very different measuring principles.
In the case of LTMFM, forces acting on the magnetic tip are detected, e.g.~allowing for the imaging of Abrikosov vortices in superconductors\cite{Moser95,Volodin02}.
For the nanoSQUID, signals caused by the entrance of such vortices are indistinguishable from signals produced by a SMP. Therefore the \textit{in situ} imaging of Abrikosov vortices is an important prerequisite for reliable nanoSQUID magnetometry.
In the first part of the manuscript, with this technique we identify the position of trapped flux in the superconducting lead of the nanoSQUID operated in high magnetic fields.
In contrast to the LTMFM, the nanoSQUID directly measures the stray flux from the magnetic tip coupled to the SQUID loop.
Therefore, in the second part, we present measurements of $\Phi(\vec{r})$ for the half space above the nanoSQUID by scanning a LTMFM with a nanoscale ferromagnet at position $\vec{r}$ as a magnetic tip.
These findings are not exclusive to the use of a Ni nanotube, but should be valid for a wide range of SMPs.
Furthermore, we show that the nanoSQUID can be used as a highly sensitive detector of displacement of the Ni nanotube.

\section{SQUID layout, properties and readout}

\begin{figure}[b]
\centering
\includegraphics[width=8cm]{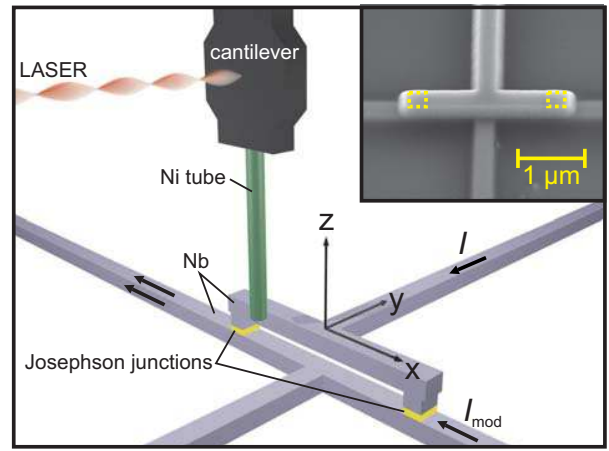}
\caption{(Color online) Schematic view (not to scale) of the nanoSQUID and Ni nanotube geometry, indicating $x,y,z$ directions as used below, with the origin centered on the surface of the upper Nb layer.
Thick arrows indicate flow of applied bias current $I$ and modulation current $I_\mathrm{mod}$.
Inset shows scanning electron microscopy (SEM) image of the Nb nanoSQUID; dotted lines indicate the two Josephson junctions.
}
\label{fig: SEM_SQUID}
\end{figure}

For the experiments presented here, we use a dc SQUID which has a sandwich-like geometry (see Fig.~\ref{fig: SEM_SQUID}), i.e., the two arms of the SQUID loop lie directly on top of each other, and are connected via two $200\times 200\,\rm{nm}^2$ planar Nb/HfTi/Nb Josephson junctions.\cite{Hagedorn02,Hagedorn06}
For this geometry, the size of the SQUID loop (in the $x-z$ plane) is given by the gap ($\sim 225\,\rm{nm}$) between top and bottom Nb layers and the lateral distance ($\sim 1.8\,\rm{\mu m}$) between the two junctions.
Using such a geometry, a very small loop size, and hence a small loop inductance of a few pH or even lower can be achieved, which is essential for obtaining very low values for $S_\Phi$.\cite{VanHarlingen82}
The rms flux noise for the SQUID used here is $S_\Phi^{1/2}\approx 220\,{\rm n}\Phi_0/\sqrt{\rm Hz}$ (in the white noise limit above $\sim 1\,$kHz).
This value was measured in a separate, magnetically and electrically shielded setup, using a sensitive cryogenic amplifier for SQUID readout.

The nanoSQUID is mounted in a vacuum chamber (pressure $\le 1\times10^{-6}\,\rm{mbar}$) at the bottom of a continuous-flow $^3$He cryostat.
The SQUID is biased at a current $I$ slightly above its critical current and at a magnetic flux $\Phi_\mathrm{mod}\propto I_\mathrm{mod}$ coupled via the modulation current $I_\mathrm{mod}$ to the loop (cf.~Fig.~\ref{fig: SEM_SQUID}).
In order to maintain operation of the SQUID at its optimum working point, i.e., at the maximum slope of its $V(\Phi)$ curve, we use a flux-locked loop (FLL) with a room-temperature voltage preamplifier.
The FLL couples a feedback flux $\Phi_f=-\Phi$ in order to compensate for any flux signal $\Phi$.
Using such a scheme, the output voltage $V_\mathrm{out}\propto I_\mathrm{mod}$ provided by the feedback loop is directly proportional to $\Phi$; in our case $V_\mathrm{out}/\Phi=2.55\,{\rm V}/\Phi_0$.

\section{LTMFM setup}

The magnetic tip used in our LTMFM setup is a $\ell=$ $6\,\mu\rm{m}$ long Ni nanotube which is fabricated by atomic layer deposition of Ni and a $\sim$25\,nm thick AlO$_x$ interlayer on a $75\,$nm diameter GaAs nanowire.\cite{Rueffer12}
The outer diameter $D_a=190(\pm35)\,\rm{nm}$ \cite{Note-1}, yielding a thickness $t=32.5(\pm17.5)\,\rm{nm}$ of the Ni layer and hence a volume of the Ni tube $V_{\rm{Ni}}=0.096(\pm0.063)\,\mu\rm{m}^3$.
The Ni nanotube is affixed parallel to the cantilever axis ($z$-axis) such that it protrudes from the cantilever end by $4\,\mu\rm{m}$.
We define the position $\vec{r}=(x,y,z)$ of the Ni tip (relative to the SQUID) as the intersection point of its cylindrical axis with the bottom end of the tube.
The cantilever hangs above the SQUID in the pendulum geometry, i.e., perpendicular to the scanned surface (in the $x$-$y$ plane; cf.~Fig.~\ref{fig: SEM_SQUID}).\cite{Gysin11}
A 3D piezo-electric positioning stage (Attocube Systems AG) moves the SQUID relative to the Ni nano-magnet.
In non-contact scanning force microscopy, the above described configuration prevents the tip of the cantilever from snapping into contact with the sample surface and thus allows for the use of particularly soft -- and therefore sensitive -- cantilevers (spring constant $\le1\,\rm{mN}/\rm{m}$).
The single-crystal Si cantilever used here is $120\,\mu\rm{m}$ long, $4\,\mu\rm{m}$ wide, and $0.1\,\mu\rm{m}$ thick and includes a $15\,\mu\rm{m}$ long, $1\,\mu\rm{m}$ thick mass on its end.\cite{Chui03}
The oscillation of the lever along $y$-direction is detected using laser light focused onto a $10\,\mu\rm{m}$ wide paddle near the mass-loaded end and reflected back into an optical fiber interferometer.\cite{Rugar89}
$100\,\rm{nW}$ of light are incident on the paddle from a temperature-tuned $1550\,\rm{nm}$ distributed feedback laser diode.

At temperature $T=4.3\,\rm{K}$ and applied magnetic field $H=0$, the nano-magnet-loaded cantilever has a resonance frequency $f_{\rm{res}}=3413\,\rm{Hz}$ and an intrinsic quality factor $Q_0=3.4\times10^4$.
Its spring constant is determined to be $k=90\,\rm{\mu N/m}$ through measurements of its thermal noise spectrum at several different temperatures.
As a result, far from the SQUID, where surface interactions do not play a role,\cite{Stipe01,Kuehn06} the cantilever has a thermally limited force sensitivity of $10\,\rm{aN}/\sqrt{\rm Hz}$.
The interferometric cantilever deflection signal is fed through a field programmable gate array (FPGA) (National Instruments) circuit back to a piezoelectric element which is mechanically coupled to the cantilever.
In this way, it is possible to self-oscillate the cantilever at its fundamental resonance frequency and at a desired amplitude.

We produce non-contact force microscopy images by scanning (in the $x$-$y$ plane for fixed $z$) the position of the nanomagnet-tipped cantilever over the SQUID and simultaneously measuring the cantilever resonance frequency $f_{res}(x,y)$, which is proportional to the force gradient $\partial F/\partial y$ acting on the nanomagnet-tipped cantilever.
From such images we can identify the topography of the nanoSQUID, allowing us to precisely position the Ni nano\-tube with respect to the nanoSQUID.
At the same time, due to the magnetization of the Ni nano\-tube tip, the images show features produced by the diamagnetic response of the superconductor.

\section{MFM imaging of an Abrikosov vortex}

\begin{figure}[t]
\centering
\includegraphics[width=8cm]{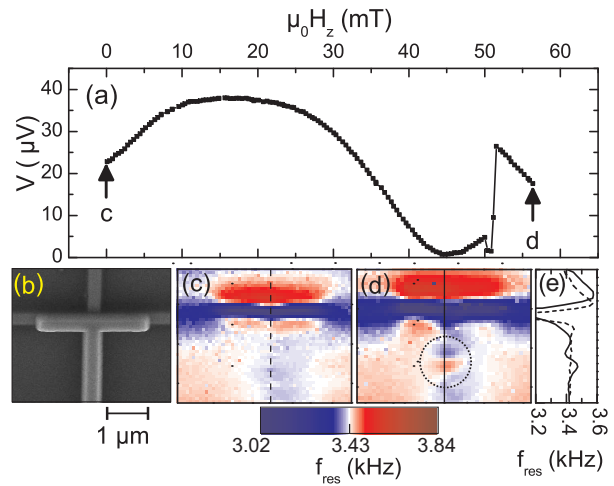}
\caption{MFM imaging of trapped flux: (a) $V(H)$ for a single sweep from 0 to $56\,\rm{mT}$; labels c, d indicate field values for LTMFM images shown in (c) and (d), respectively.
(b) SEM image of the nanoSQUID.
(c,d) LTMFM images $f_{res}(x,y)$, without trapped vortices at $H=0$ (c) and with a trapped vortex (indicated by dotted circle) at $\mu_0H=56\,\rm{mT}$ (d).
(e) linescans along dashed line in (c) (dashed curve) and solid line in (d) (solid curve).
}
\label{Fig:SQUID_flux}
\end{figure}

Prior to the measurements of the magnetic flux coupled by the Ni tip to the SQUID, we investigate a possible impact of an applied magnetic field on the nanoSQUID, since Abrikosov vortices may enter the superconducting areas, degrading its performance.
$H$ is aligned along $z$ direction, with a possible tilt of a few degrees.
The trapping of a vortex appears as a voltage jump in the periodic $V(H)$ characteristics, when the SQUID voltage is measured directly, rather than using the FLL readout.
Note that the SQUID voltage oscillates with increasing $H$, due to the non-perfect alignment of $H$ along the $z$ direction, i.e., $H$ has a finite in-plane component, which induces magnetic flux threading the SQUID loop.
From the effective area of the SQUID and the oscillation period of $V(H)$, we estimate a tilt of the applied field of $\sim 2\,^\circ$ with respect to the $z$ axis.
An example for a vortex trapping process is shown in Fig.~\ref{Fig:SQUID_flux}(a), where a huge jump in $V(H)$ occurs near $\mu_0H=50\,\rm{mT}$.
We note that during the field sweep, the Ni tip was retracted from the SQUID.
For further improvement of the SQUID layout the knowledge of the position of trapped vortices is indispensable.
The ability of the LTMFM setup to image stray fields can be used to visualize vortices in the superconductors as well as the magnetic field of the screening currents of the nanoSQUID itself.
In Fig.~\ref{Fig:SQUID_flux}(d), taken at a magnetic field above the jump in $V(H)$, such a vortex is visible in the superconducting lead (top Nb layer) of the nanoSQUID.
The vortex appears as a distortion in the otherwise flat resonance frequency $f_{\rm{res}}$ distribution along the Nb-line (cf.~linescans in Fig.~\ref{Fig:SQUID_flux}(e)). 
In the given setup $\Delta f_{\rm{res}}\propto\partial^2 B_z/\partial y^2$, i.e., a tripolar response is expected, which might be distorted due to the influence of the screening currents in the SQUID or a remaining magnetization of the Ni nano\-tube in the $x$-$y$-plane.
In contrast, at fields below the jump the trapped vortex is absent [see Fig.~\ref{Fig:SQUID_flux}(c)].
For the subsequent investigations, we operate the SQUID in nominally zero field only, i.e., trapped vortices do not play a role.

\section{Experimental determination of $\Phi(\vec{r})$}

In order to determine $\Phi(\vec{r})$ we measure the nanoSQUID signal, i.e., the magnetic flux $\Phi$ through the SQUID loop as a function of the Ni nano\-tube position $(x,y)$ for fixed $z$.
Such measurements produce images $\Phi(x,y)$ of the spatially dependent magnetic coupling of the Ni nano\-tube to the nanoSQUID.
The experiment was performed in the following way:
First, we bring the Ni nano\-tube into a well-defined saturated magnetic state along its easy axis.
This is done by a half magnetization cycle, i.e., a sweep of $H$ (aligned along $z$ direction, as above) from zero to $\mu_0H_\mathrm{max}=-150\,\mathrm{mT}$ and back to $H=0$.
From previous experiments, we know that $H_\mathrm{max}$ is strong enough to saturate the magnetization of the Ni nano\-tube.\cite{Rueffer12,Weber12}
In order to avoid trapped flux in the SQUID, the magnetization cycle is performed at $T=14\,$K, i.e., significantly above the transition temperature $T_c\sim 9\,$K of the Nb SQUID.
Subsequently, we zero-field cool the SQUID to its operation temperature $T=4.3\,$K, and then set up the FLL readout for the SQUID.
For various distances $z$ between the tip and the top Nb layer of the SQUID we make scans in the $x$-$y$ plane with a scan range of about $6 \times 7 \,\mathrm{\mu m}^2$ corresponding to $81\times81$ pixels.
The scans start at the largest distance of $z\approx700\,\mathrm{nm}$.
In steps of $50\,\mathrm{nm}$ the distance is subsequently reduced until the tip touches the top Nb layer of the SQUID (at $z=0$), which is detected as a loss of the oscillation of the cantilever.
The touchpoint is also necessary for the calibration of the $z=0$ position.

\begin{figure*}[t]
\centering
\includegraphics[width=12.705cm]{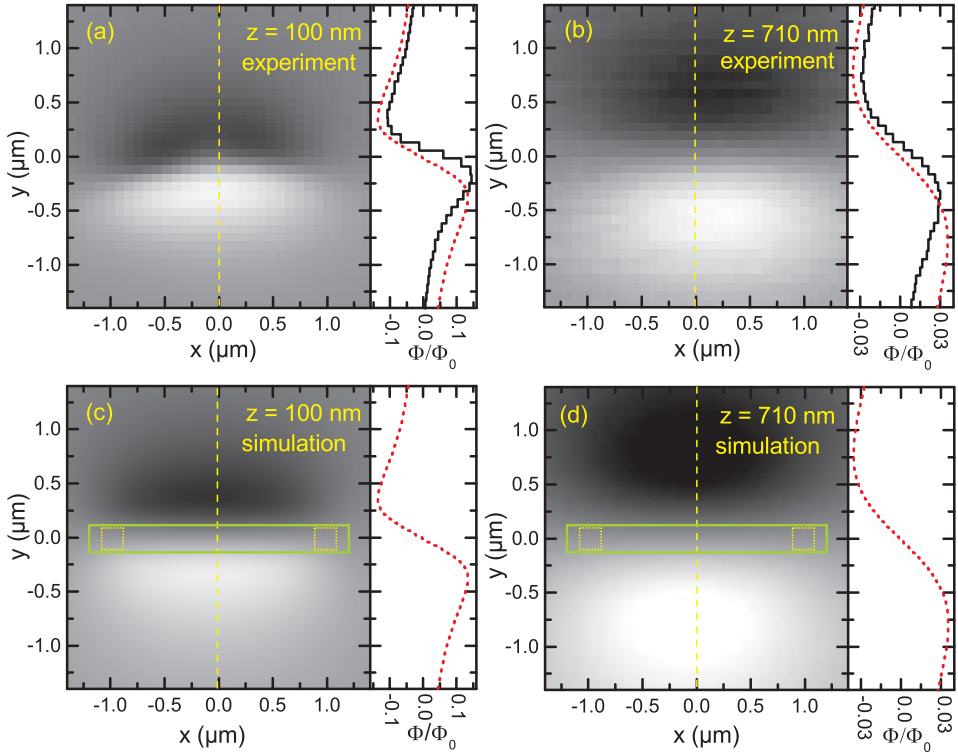}
\caption{Magnetic flux $\Phi$ generated in the SQUID vs $x$-$y$ position of the Ni nano\-tube (magnetized along $z$-axis).
In (c) and (d), solid rectangle and dotted squares indicate position of the SQUID and the two junctions, respectively.
Vertical dashed lines indicate position of linescans $\Phi(y)$ to the right of each image.
Upper graphs (a, b) show experimental results and lower graphs (c, d) show corresponding simulation results for $z=100\,$nm (left graphs) and $z=710\,$nm (right graphs).
For the simulation we assumed $V_{\rm{eff}}=0.047\,\mu{\rm m}^3$.
Linescans in (a,b) also include calculated linescans from (c,d).
}
\label{Fig:Coupling}
\end{figure*}

Figure \ref{Fig:Coupling}(a,b) shows two representative $\Phi(x,y)$ images taken at $z=100\,\mathrm{nm}$ (a) and $z= 710\,\mathrm{nm}$ (b).
The images show a bipolar flux response, i.e., when the tip crosses the SQUID loop the flux signal is inverted.
For the closer distance [Fig.~\ref{Fig:Coupling}(a)] the induced flux is stronger and spatially more confined as compared with the larger distance [Fig.~\ref{Fig:Coupling}(b)].
At $z=100\,\mathrm{nm}$, we obtain $\Delta \Phi=\Phi_\mathrm{max}-\Phi_\mathrm{min}\approx 0.26\,\Phi_0$, with the positions of the maximum $\Phi_\mathrm{max}$ and minimum $\Phi_\mathrm{min}$ in the linescan $\Phi(y)$ (at $x=0$) being separated by $\Delta y=370\,$nm.
For $z=710\,\mathrm{nm}$, we find $\Delta \Phi \approx 0.06\Phi_0$ and $\Delta y=750\,$nm.

\section{Analysis with $\Phi_\mu(\vec{r})$}

\begin{figure}[b]
\centering
\includegraphics[width=8cm]{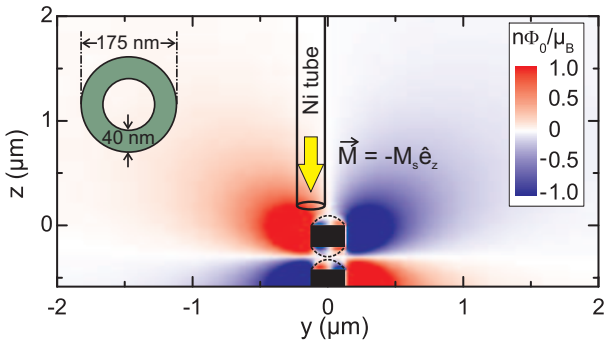}
\caption{(color online) Calculated coupling factor $\phi_\mu$ in the $y$-$z$ plane ($x=0$) for a point-like magnetic particle with magnetic moment $\vec{\mu}$ along $-\hat{e}_z$.
Black rectangles indicate position of the Nb top and bottom layer; dotted lines include regions for which the simulations produce unphysical results.\cite{note3}
A sketch of the bottom part of the Ni nano\-tube (drawn to scale) is shown within the coupling map in order to illustrate the spatial dependence of the coupling factor within the volume of the tube.
Upper left inset schematically shows a zoomed cross-section of the Ni nano\-tube.
}
\label{Fig:Niwire_Coupling}
\end{figure}

In order to analyze the measured flux signals, we start from numerical simulations of $\phi_\mu(\vec{r}_p)$ for a point-like SMP at position $\vec{r}_p$.
Figure \ref{Fig:Niwire_Coupling} shows the calculated coupling factor $\phi_\mu$ in the $y$-$z$ plane, with the SQUID loop in the $x$-$z$ plane and the magnetic moment pointing along the $-z$-direction.
$\phi_\mu$ decreases with increasing distance from the SQUID loop and inverts when crossing the SQUID loop.
This spatial dependence has a strong impact on the magnetic flux $\Phi(\vec{r})$ which is coupled by a Ni nano\-tube (at position $\vec{r}$) with finite size into the SQUID.
For the calculation of $\Phi(\vec{r}$) we integrate $\phi_\mu$ over the volume $V_\mathrm{Ni}$ of the Ni nano\-tube at position $\vec{r}$ and multiply this with the Ni saturation magnetization $M_s$, i.e., $\Phi(\vec{r})=M_s\int_{V_\mathrm{Ni}(\vec{r})}{\phi_\mu(\vec{r}_p)dV}$, assuming a homogeneous $M_s$ over the entire volume of the Ni nano\-tube.

Figure \ref{Fig:Coupling} (c) and (d) show calculated flux images $\Phi(x,y)$ for $z=100$ and 710\,nm, respectively, of a tube with the above mentioned geometric dimensions and a saturation magnetization $M_s$ along $-z$-direction [cf.~Fig.\ref{Fig:Niwire_Coupling}].
The bipolar flux response and the positions of the minima $\Phi_\mathrm{min}$ and maxima  $\Phi_\mathrm{max}$ in $\Phi(y)$ (for $x=0$) are reproduced well by the simulations.

\begin{figure}[t]
\centering
\includegraphics[width=8cm]{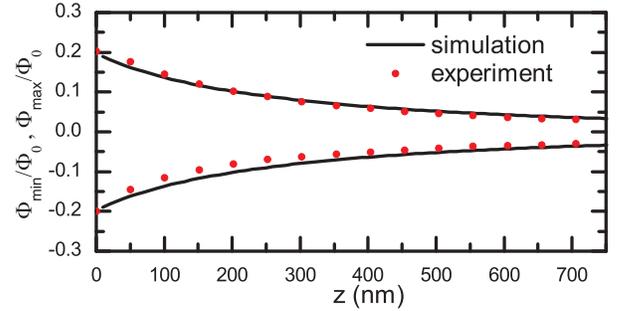}
\caption{Experimental and simulated minimum and maximum flux signals, $\Phi_\mathrm{min}$ and $\Phi_\mathrm{max}$, versus distance $z$.
For the simulation we assumed $V_{\rm{eff}}=0.047\,\mu{\rm m}^3$.
}
\label{Fig:Coupling2}
\end{figure}

For a quantitative analysis, Fig.~\ref{Fig:Coupling2} compares experimentally obtained $\Phi_\mathrm{min}$ and $\Phi_\mathrm{max}$ for all investigated distances to the simulated ones.
From previous work on similar Ni nano\-tubes \cite{Rueffer12,Weber12}, we know that the saturation magnetization is equal within the error to the bulk value known from the literature $M_s=408\,$kA/m \cite{Kittel}. 
Assuming this saturation magnetization for the nano\-tubes used here, we find that an effective volume of the Ni-shell $V_{\rm{eff}}=0.047\,\mu{\rm m}^3$ gives the best quantitative agreement with the data.  
This value for $V_{\rm{eff}}$ is significantly smaller than the value for $V_{\rm{Ni}}$ as quoted above, but still within the large uncertainty for $V_{\rm{Ni}}$. 
Furthermore, an independent determination of $V_{\rm{eff}}$ on the same Ni tube, via cantilever magnetometry yields a value which is even somewhat below the one obtained via SQUID measurement.
We note, that the formation of a multidomain state close to the bottom of the Ni tube is unlikely, as hysteresis curves $M(H)$ measured both with SQUID and cantilever magnetometry indicate no reduction of magnetic signal upon sweeping $H$ from $\pm H_\mathrm{max}$ back to zero.
In the experiment we also find an asymmetry in $\Phi(y)$, i.e. $\Phi_\mathrm{max}\ge |\Phi_\mathrm{min}|$.
This effect is most likely caused by flux focusing effects of the feed lines in the top and bottom Nb layers, which are not considered in the simulations.
The flux focusing effects are also visible in the distorted flux image (broken horizontal symmetry) in Fig.\ref{Fig:Coupling}(a).
An additional asymmetry may be caused by a slightly tilted tube with respect to the SQUID plane.
In our case however, this effect is considered to be small, since the tilt angle is measured to be less than $5^\circ$.

\section{Displacement detection}

Finally, we discuss the sensitivity of our setup for the detection of the oscillatory motion of the cantilever by the SQUID.\cite{Usenko11,Vinante11}
While the absolute flux signal from the Ni nano\-tube is optimally detected at the positions yielding $\Phi_\mathrm{max}$ and $\Phi_\mathrm{min}$, for the cantilever displacement detection, a large gradient $\partial\Phi/\partial y$ is required.
The linescans in Fig.~\ref{Fig:Coupling} clearly show that the optimum position for displacement detection is directly above the SQUID.
For our device, we find for $z=50\,\mathrm{nm}$ a gradient $\Phi_y\equiv\partial\Phi/\partial y=2\times10^6\,\Phi_0/\rm{m}$.
With the flux noise $S_{\Phi}^{1/2}=220\,\rm{n}\Phi_0/\sqrt{\rm Hz}$, this yields an extremely low value for the predicted displacement sensitivity $S_{r}^{1/2}= S_{\Phi}^{1/2}/\Phi_y = 110\,\mathrm{fm}/\sqrt{\rm Hz}$, which is already a factor of two below the best value found in literature.\cite{Usenko11, Vinante11, Vinante12}
Still, $S_r$ is by far not optimized and could be further improved by using a reduced linewidth for the SQUID arm in the top Nb layer and by increasing the number of spins in the magnet.

\section{Conclusions}

In conclusion, we experimentally determined the spatial dependence of the magnetic coupling between a Ni nano\-tube and a Nb nanoSQUID.
Operating the nanoSQUID in a flux locked loop, we measured the flux through the SQUID loop $\Phi(\vec{r}$) generated by the Ni nano\-tube during the scan of the tip in 3D space above the nanoSQUID.
Our results are in good agreement with a recently developed routine for numerical calculation of the coupling factor between a small magnetic particle and a nanoSQUID.
This provides an important step toward the development of optimized nanoSQUIDs for the investigation of small magnetic particles.
With the presented measurement system, we demonstrate a reliable and non-destructive \textit{in situ} tool for the challenging task of positioning a nano-scaled magnet to the position of highest coupling of a nanoSQUID.
Furthermore, with a proper readout technique, our highly flux-sensitive nanoSQUID can be used for displacement detection of the cantilever in an MFM with extremely good displacement sensitivity, which still can be further improved.
By using an MFM imaging mode, we also demonstrate the imaging of Abrikosov vortices, which are trapped at high magnetic fields in the superconducting leads of the nanoSQUID. 
This technique is useful for the improvement of the high-field suitability of nanoSQUIDs as well as for the differentiation between a signal originated by a SMP or by the entrance of a spurious Abrikosov vortex.
Finally, we demonstrate the use of a nanoSQUID as a local probe of the stray fields produced by the Ni nano\-tube, which may be of great importance in understanding magnetization reversal in these magnetic nanostructures. Such investigations will be the subject of future work as will investigations of other SMPs.

\begin{acknowledgements}

J.~Nagel acknowledges support by the Carl-Zeiss-Stiftung.
This work was funded by the Deutsche Forschungsgemeinschaft (DFG) via the SFB/TRR 21, by the European Research Council via ERC advanced grant SOCATHES, by the Canton Aargau, and by the Swiss National Science Foundation (SNF, Grant No.~200020-140478).
MP, ER, and AFiM acknowledge support from the NCCR QSIT.
The research leading to these results has received funding from the European Community's Seventh Framework Programme (FP7/2007-2013) under Grant Agreement No.~228673 MAGNONICS.

\end{acknowledgements}

\providecommand*\mcitethebibliography{\thebibliography}
\csname @ifundefined\endcsname{endmcitethebibliography}
  {\let\endmcitethebibliography\endthebibliography}{}


\begin{mcitethebibliography}{54}
\bibitem[Not()]{Note-1}
The determination of $D_a$ for the tube investigated here was not possible, as
  the tube was lost during warm up after our measurements. Instead, we
  determined $D_a$ for 6 other tubes from the same batch by SEM. Those had
  values of $D_a$ varying from $155-225\,$nm with a variation in $D_a$ of
  $10-20\,$nm along a single tube.\relax
\mciteBstWouldAddEndPunctfalse
\mciteSetBstMidEndSepPunct{\mcitedefaultmidpunct}
{}{\mcitedefaultseppunct}\relax
\EndOfBibitem
\bibitem[not()]{note3}
For simulating the current distribution, we have to assume that the currents are flowing in surface sheets rather than across the entire width of the two Nb lines (in y-direction). This assumption allows for coupling of flux between the surface sheets, which produces the localized minima and maxima very close to the two SQUID arms, as visible in Fig.~\ref{Fig:Niwire_Coupling}.\relax
\mciteBstWouldAddEndPunctfalse
\mciteSetBstMidEndSepPunct{\mcitedefaultmidpunct}
{}{\mcitedefaultseppunct}\relax
\EndOfBibitem
\end{mcitethebibliography}
\end{document}